\documentclass{PoS}
\usepackage{subfig}
\usepackage{ulem}

\title{15 GHz VLBI detection of the HST-1 feature in the M\,87 jet}

\ShortTitle{15 GHz VLBI detection of the HST-1 feature in the M\,87 jet}

\author{\speaker{C.S.~Chang}$^{,a,}$\thanks{Member of the International Max Planck Research School for Astronomy and Astrophysics}\,,  E.~Ros$^a$, Y.~Y.~Kovalev$^{a,b,}$\thanks{Alexander von Humboldt Research fellow}\,, M.~L.~Lister$^c$ \\
      \llap{$^a$}Max-Planck-Institut f\"ur Radioastronomie, Auf dem H\"ugel 69, D-53121 Bonn, Germany\\
      \llap{$^b$}Astro Space Centre of Lebedev Physical Institute, Profsoyuznaya 84/32, 117997 Moscow, Russia\\
      \llap{$^c$}Department of Physics, Purdue University, 525 Northwestern Avenue, West Lafayette, IN 47907, USA\\
        E-mail: \email{cschang@mpifr.de},
        \email{ros@mpifr.de},
        \email{ykovalev@mpifr.de},
        \email{mlister@physics.purdue.edu}
        }

\abstract{A bright feature 100 pc away from the core in the powerful jet of M\,87 shows mysterious properties. Earlier radio, optical and X-ray observations have shown that this feature, labelled HST-1, is superluminal, and is possibly connected with the TeV flare detected by HESS in 2005. To examine the possible blazar-like nature of HST-1, we analyzed $\lambda$2\,cm VLBA data from dedicated full-track observations and the 2\,cm survey/MOJAVE VLBI monitoring programs observed from 2000 to 2008. Applying wide-field imaging techniques, the HST-1 region was imaged
at milliarcsecond resolutions. Here we present the first 15\,GHz VLBI detection of this feature and discuss the connection between our radio findings and the TeV detection.}

\FullConference{The 9th European VLBI Network Symposium on The role of VLBI in the
Golden Age for Radio Astronomy and EVN Users Meeting\\
                 September 23-26, 2008\\
                 Bologna, Italy}

\begin{document}

\section{Introduction}
Active Galactic Nuclei (AGN) are among the most interesting phenomena in the Universe, and they have been studied since the first discovery in 1943 \cite{seyfert43}. Until now, although there are clues that imply a super-massive black hole (SMBH) is the engine launching the powerful jet \cite{urry95}, the exact mechanism remains unknown. 

M\,87 (also known as Virgo\,A) is a nearby elliptical galaxy located in the Virgo cluster. It hosts a very powerful one-sided jet emerging from the central region. Observations show that M\,87 contains $2.4\times10^{9}$ $M_{\odot}$ within a 0.25$^{\prime\prime}$ radius, which give hints of a SMBH in the center \cite{harms94}. Due to its closeness (16\,Mpc), M\,87 is a major candidate for studying AGN phenomena, and has been monitored at different wavelengths over the last decades. Superluminal motion was reported from {\it Hubble Space Telescope} ({\it HST\/}) observations within 6$^{\prime\prime}$ of the jet with speeds of 4\,$c$ to 6\,$c$ \cite{biretta99}. Discrepant speeds were reported with values between 0.25\,$c$ to 0.4\,$c$ \cite{ly07}, and a value of 2\,$c$ \cite{walker08} from VLBA 43\,GHz observations. VLBA $\lambda$2\,cm observations found apparent speeds <\,0.4\,$c$ during 1994 to 2001 \cite{kellermann04,kovalev07}. Therefore, the kinematical properties of the jet in M\,87 are still under discussion.  From the \textit{HST} observation in 1999, a bright knot along the jet was discovered, and was named HST-1. This feature is located 0.8$^{\prime\prime}$ away from the core, which corresponds to a projected distance of 0.1\,kpc, and is active in the radio, optical and X-ray regimes. VLBA $\lambda$20 cm observations show that HST-1 has sub-structure and contains superluminal moving components up to 4\,$c$ \cite{cheung07}. These observations interpret HST-1 as a collimated shock in an AGN jet.

However, recent multi-wavelength observations show that HST-1 could be related to the origin of the TeV emission of M\,87. In 2005, the HESS telescope detected a TeV flare from M\,87 \cite{aharonian06}. Comparing with the results in soft X-rays (\textit{Chandra}), and VLA $\lambda$2\,cm observations \cite{cheung07, harris06}, the light curves of HST-1 reach the maximum in 2005, while the resolved core shows no correlation with the TeV flare. Therefore, the TeV emission from M\,87 might be associated to HST-1 \cite{harris08}. The AGN standard model considers the blazar behavior to originate at the vicinity of the SMBH. However, HST-1 is 100\,pc away from the core. If the hypothesis is true, the present AGN model would face a challenge. Here we examined this hypothesis with $\lambda$2\,cm VLBI wide-field imaging of the HST-1 feature.

\section{Data and Wide-field Imaging}
M\,87 has been monitored at $\lambda$2\,cm with the VLBA since 1994 by the 2\,cm Survey/MOJAVE programs\footnote{\tt http://www.cv.nrao.edu/2cmsurvey/, http://www.physics.purdue.edu/MOJAVE/} \cite{kellermann04, lister05}. We have analyzed ten epochs of this monitoring program after late 2001, together with three observing sets of targeted observations on M\,87 in 2000--2001 (see Table \ref{tab:epoch_list}). 

\begin{table}
\centering
\begin{tabular*}{0.635\textwidth}{@{}llccl@{}}
\hline
\hline
 & Exp.\ & $t_\mathrm{int}$$^\mathrm{a}$ & rms    &  \\
Epoch & Code  & \footnotesize{[min]}          & \footnotesize{[mJy\,beam$^{-1}$]} & HST-1 \\
\hline
2000.06                & \texttt{BK073A}$^\mathrm{b,c}$  & 476 & 0.080 & Not detected \\
2000.35                & \texttt{BK073B}$^\mathrm{b,c}$  & 476 & 0.081 & Not detected \\
2000.99                & \texttt{BK073C}$^\mathrm{b,c}$  & 476 & 0.12 & Not detected \\
2001.99                & \texttt{BR077D}$^\mathrm{d}$    & 68  & 0.23 & Not detected \\
2002.25                & \texttt{BR077J}$^\mathrm{d}$    & 57  & 0.25 & Not detected  \\
2004.61                & \texttt{BL111N}$^\mathrm{d}$    & 63  & 0.20 & Detected \\
2004.92                & \texttt{BL111Q}$^\mathrm{d}$    & 64  & 0.29 & Detected \\
2005.30$^\mathrm{e}$   & \texttt{BL123E}$^\mathrm{d}$    & 63  & 0.16 & Detected \\
2005.85$^\mathrm{e}$   & \texttt{BL123P}$^\mathrm{d}$    & 63  & 0.23 & Detected \\
2006.45                & \texttt{BL137F}$^\mathrm{d}$    & 22  & 0.33 & Not detected \\
2007.10                & \texttt{BL137N}$^\mathrm{d}$    & 42  & 0.36 & Not detected \\
2007.42                & \texttt{BL149AA}$^\mathrm{d}$   & 45  & 0.38 & Not detected \\
2008.33                & \texttt{BL149AO}$^\mathrm{b}$   & 41  & 0.27 & Not detected \\
\hline
\multicolumn{5}{@{}l@{}}{\footnotesize{$^\mathrm{a}$ Total scheduled VLBA on-source time}} \\
\multicolumn{5}{@{}l@{}}{\footnotesize{$^\mathrm{b}$ Sampling rate: 256\,Mbit\,s$^{-1}$}} \\
\multicolumn{5}{@{}l@{}}{\footnotesize{$^\mathrm{c}$ Dedicated full-track experiment on M\,87 (see  \cite{kovalev07}); the array includes a }}\\
\multicolumn{5}{@{}l@{}}{\footnotesize{\,\,\,\,single VLA antenna (Y1)}} \\
\multicolumn{5}{@{}l@{}}{\footnotesize{$^\mathrm{d}$ Sampling rate: 128\,Mbit\,s$^{-1}$}} \\
\multicolumn{5}{@{}l@{}}{\footnotesize{$^\mathrm{e}$ Images shown at the upper (2005.30) and lower panel (2005.85) in Fig. \ref{fig:HST-1 overlay}}}\\
\end{tabular*}
\caption{Journal of VLBA $\lambda$2\,cm observations of M\,87}
\label{tab:epoch_list}
\end{table}

Before imaging HST-1, two issues needed to be considered. 
First, HST-1 lies 800 beam sizes away from the brightest feature (the VLBA core). For this reason, averaging the data would produce time and bandwidth smearing in the HST-1 region. Second, based on earlier VLBA $\lambda$20\,cm and VLA $\lambda$2\,cm observations of HST-1 \cite{cheung07}, we expect the total 15\,GHz flux density to be at milli-Jansky level, which is much weaker than the total flux density of the inner-jet ($\sim$2.5\,Jy). In order to detect HST-1, we need to image the inner-jet region with its extended structure (over tens of milliarcseconds); otherwise, the sidelobes from the core would cover the HST-1 emission. To reach this goal, we applied natural weighting and tapering with a Gaussian factor of 0.3 at a radius of 200 M$\lambda$ in the $(u,v)$-plane to weigh down the long baselines on all datasets. The resultant resolution of our images was of $\sim$2$\times$1\,milliarcseconds (mas) at a position angle of $\sim$13$^{\circ}$.
The extended inner-jet \textsc{clean} model was applied to the un-averaged data with \textsc{calib} in \textsc{aips}. When the inner-jet model was well-established, we used \textsc{imagr} to define a second cleaning field of the HST-1 region, and proceeded with further loops of \textsc{clean}ing and phase and amplitude self-calibration. By applying wide-field imaging techniques to the data, we obtained an image of HST-1 with the highest resolution. 

\section{Results and Discussion}
\begin{figure}
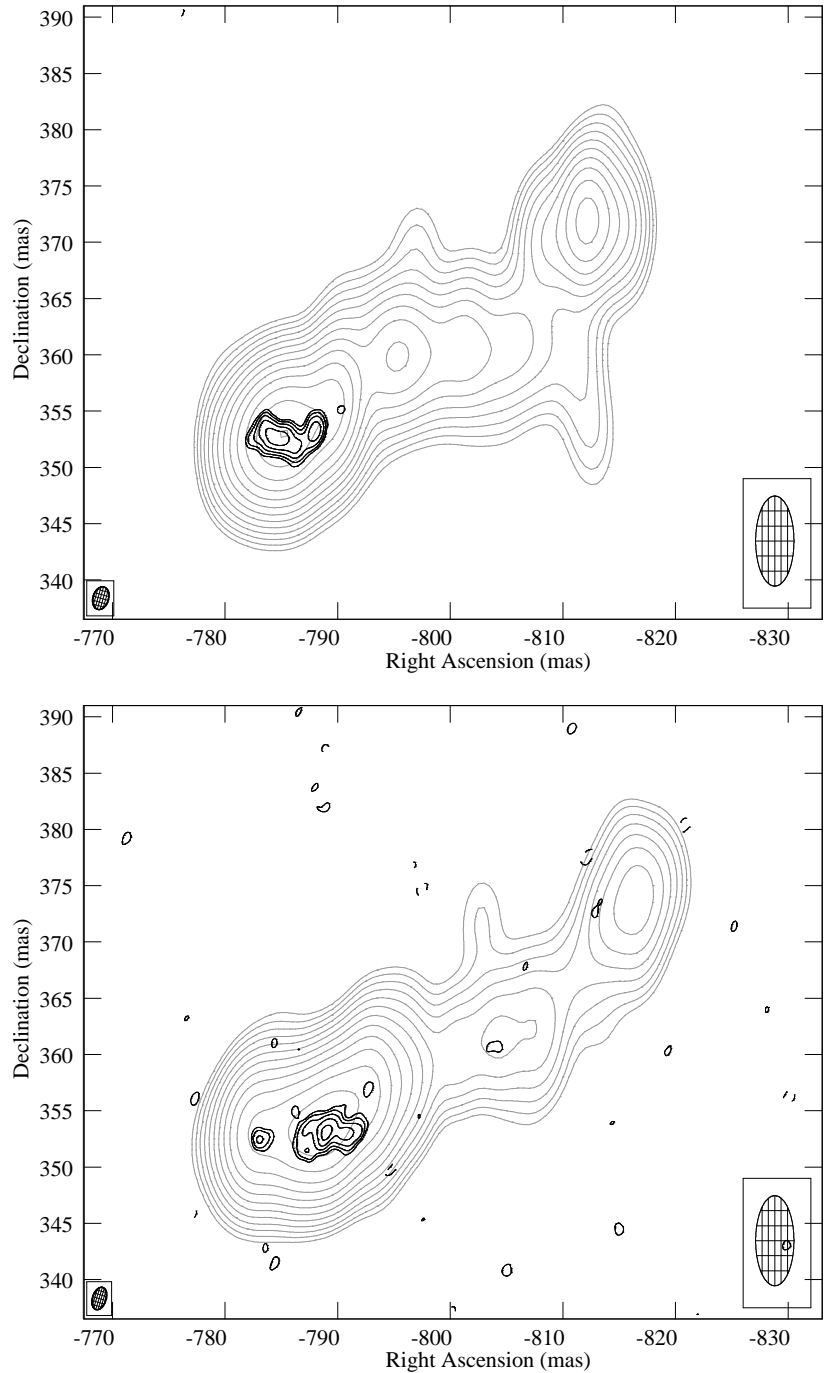

\centering
\includegraphics[angle=270, width=0.75\textwidth]{2005.3_overlay_mas_final.ps}
\includegraphics[angle=270, width=0.75\textwidth]{2005.9_overlay_mas_final.ps}
\caption{VLBA images of the HST-1 region in M\,87 at $\lambda$\,2\,cm (black contour, beam size at half power level\,=\,2$\times$1\,mas, P.A.\,=\,$-16^{\circ}$, plotted bottom left) and $\lambda$\,20\,cm (grey contour, beam size\,=\,8$\times$3\,mas, P.A.\,=\,0$^\circ$, plotted bottom right). Upper panel: epoch 2005.30 ($\lambda$\,2\,cm, peak surface brightness: 3.4\,mJy\,beam$^{-1}$) and 2005.35 ($\lambda$\,20\,cm, peak: 45\,mJy\,beam$^{-1}$); lower panel: epoch 2005.85 ($\lambda$\,2\,cm, peak: 3.3\,mJy\,beam$^{-1}$) and 2005.82 ($\lambda$\,20\,cm, peak: 42\,mJy\,beam$^{-1}$). The coordinates are relative to the core location. The lowest contour is 0.7\,mJy\,beam$^{-1}$, and the contour levels are separated by a factor of $\sqrt2$.}
\label{fig:HST-1 overlay}
\end{figure}

With tapering and natural weighting to the data, the most extended inner-jet structures were traced until 100\,mas (8\,pc), and the noise levels of HST-1 \textsc{clean} images are listed in Table \ref{tab:epoch_list}.
The inner-jet structure had a total flux density of 2.5\,Jy, with overall flux density changes up to 0.5\,Jy during 2000 to 2008. Out of 13 epochs, HST-1 was detected in 4 epochs from 2004.61 to 2005.85 (see Table \ref{tab:epoch_list} and Figure \ref{fig:HST-1 overlay}). The feature had a peak surface brightness of 3 to 5\,mJy\,beam$^{-1}$, and the total flux density was 14 to 22\,mJy, with an extended structure of 10\,mas. These detections show that HST-1 has a weak peak at $\lambda$2\,cm. 

Due to its low flux density and extended structure, only the brightest feature could be detected. The peak position changes show an outward motion from the core. By linear-fitting the distance of the peak position to the central core as a function of time, a proper motion of 0.44\,$\pm$\,2.24 mas\,yr$^{-1}$ was derived, which corresponds to a projected linear speed of 0.11\,$\pm$\,0.56\,$c$. Previous observations report higher speeds up to 6\,$c$ from the HST-1 region \cite{biretta99, cheung07}. 
Figure \ref{fig:HST-1 overlay} shows two overlaid images of HST-1 at $\lambda$20\,cm (Cheung, priv.\ comm.) and $\lambda$2\,cm. Note that the feature detected at $\lambda$2\,cm was located at the same position as the HST-1d component by Cheung et al \cite{cheung07}. The HST-1d component shows a stationary behavior with a speed of <\,0.25\,$c$, which is comparable with our measurements.

The emission intensity of HST-1 has reached the maximum in different wavebands at 2005, and the light curves from the VLA $\lambda$2\,cm, VLBA $\lambda$20\,cm, and X-ray observations share the same tendency \cite{cheung07, harris06}. Our results were consistent with those observations. The HESS team reports that a TeV flare was detected in 2005 \cite{aharonian06}. It is believed that the very high energy emissions come from the vicinity of the SMBH, however, {\it Chandra\/} observations show that the resolved core has a flat light curve during this time, while HST-1 has a significant peak in 2005. Our results also show that the total flux density of the inner-jet does not have a peak in 2005, while HST-1 was only detected during 2004 to 2005. Those findings support that HST-1, a feature 100\,pc away from the SMBH, could be the source of the TeV emission. However, according to our detection, HST-1  appeared to be extended, which does not support the blazar type activity. Moreover, no emerging or rapidly moving features were found among the detections. Therefore, our results does not support that HST-1 was the origin of the TeV emission, however, one could not exclude the possibility.

\paragraph*{Acknowledgments}
\begin{small}
We thank M.~A.~Garrett, R.~C.~Walker, G.~Cim\`o, A.~Mor\'e, S.~M\"uhle, and C.~M.~Fromm for valuable comments and inspiring discussions. Special thanks are due to C.~C.~Cheung for providing the VLBA $\lambda$20\,cm images \cite{cheung07}. This research was supported by the EU Framework 6 Marie Curie Early Stage Training programme under contract number MEST/CT/2005/19669 ``ESTRELA''. This research has made use of data from the 2\,cm\,Survey \cite{kellermann04} and MOJAVE \cite{lister05} programs. The Very Long Baseline Array is operated by the USA National Radio Astronomy Observatory, which is a facility of the USA National Science Foundation operated under cooperative agreement by Associated Universities, Inc.
\end{small}


\end{document}